\documentclass[english,aps,prd,onecolumn]{revtex4-1}

\usepackage{bm}
\usepackage{appendix}
\usepackage{graphicx}
\usepackage{amsmath}
\usepackage{amssymb}%
\usepackage{color}
\usepackage{lipsum}
\usepackage[colorlinks=true,citecolor=blue,urlcolor=blue]{hyperref}
\usepackage[utf8]{inputenc}  
\usepackage[T1]{fontenc}     
\DeclareUnicodeCharacter{039B}{$$ \Lambda $$}

\begin{document}

\title{ Absence of charged pion condensation in a magnetic field with parallel rotation}

\author{Puyuan Bai and Lianyi He}
\affiliation{Department of Physics, Tsinghua University, Beijing 100084, China}

\date{\today}%

\begin{abstract}
We investigate the critical temperature of a relativistic Bose-Einstein condensate of charged bosons driven by rotation in a parallel magnetic field [\textcolor{blue}{Y. Liu and I. Zahed, Phys. Rev. Lett. {\bf120}, 032001 (2018)}]. 
For non-interacting bosons, the critical temperature can only be determined for a system with fixed angular momentum.  We find that the critical temperature of the non-interacting system vanishes due to the fact that the system
is quasi-one-dimensional, indicating that non-interacting bosons cannot undergo Bose-Einstein condensation. For interacting bosons, we investigate a system with quartic self-interaction.  
We show that the order parameter vanishes and the off-diagonal long-range order is absent at any nonzero temperature because of the quasi-one-dimensional feature, in accordance with the Coleman-Mermin-Wagner-Hohenberg theorem.
\end{abstract}

\maketitle

\section{Introduction}

It is expected that in noncentral nucleus-nucleus collisions,  large vorticity and magnetic fields can be generated. Theoretical studies predicted that noncentral collisions involve large angular momenta in the range $10^3\sim10^5\hbar$~\cite{PRC77:024906, EPJC75:406, PRC93:064907, PRC94:044910}.  The global polarization of $\Lambda$ hyperon observed in off-central Au-Au collisions reported by the STAR Collaboration indicates a large vorticity with an angular velocity   $\Omega \approx (9\pm1)\times10^{21}$Hz $\sim0.05m_\pi$~\cite{NAT548:62}. Meanwhile, it is expected that a large magnetic field $B$, parallel to the angular velocity, is formed at the early stage of the collision. Numerical simulations indicate that the strength of the magnetic field reaches $eB\sim m_\pi^2$~\cite{IJMPA24:5925, PRC85:044907, PPNP88:1}.  

Motivated by the large angular velocity and magnetic field created in noncentral nucleus-nucleus collisions, the state of matter under the circumstance of parallel rotation and magnetic field (PRM) arises as an interesting theoretical issue. 
Pions are the lightest hadrons of the strong interaction and are regarded as the pseudo-Goldstone bosons associated with the dynamical chiral symmetry breaking. As bosons, they may undergo Bose-Einstein condensation (BEC) in certain circumstances. The studies of quantum chromodynamics (QCD) at finite isospin chemical potential indicate that BEC of charged pions takes place when the isospin 
chemical potential exceeds the mass of charged pions \cite{PRL86:592, PRD66:034505, PRD97:054518}. It was proposed that BEC of pions may be formed in compact stars~\cite{PRL29:382, PRL29:386, PRL30:1340, PRL31:257, PRD98:094510}, in heavy ion collisions~\cite{ZPA277:391, PRL43:1705, PLB316:226}, and in the early Universe~\cite{PRL121:201302, PRL126:012701, PRD104:054007}. It was proposed that PRM provides a new mechanism for BEC of charged pions~\cite{PRL120:032001}. The argument is based on the solution of the Klein-Gordon equation for \emph{noninteracting} pions in PRM. The solution of the Klein-Gordon equation indicates that the Landau level degeneracy of charged pions in a constant magnetic field is lifted by rotation, and the rotation then plays the role of a chemical potential, which makes the pions condense.  Later numerical calculations based on variational method including pion-pion quartic interaction show that the ground state of the BEC is a giant quantum vortex where pions condense into a state with a large angular quantum number~\cite{PRD106:094010}.

As a Bose-Einstein condensate, the critical temperature is important for its realization. If the critical temperature is lower than the freeze-out temperature, the charged pion BEC cannot be formed in noncentral nucleus-nucleus collisions.  However, for the charged pion BEC driven by PRM proposed in~\cite{PRL120:032001}, the critical temperature has not yet been calculated. In this work, we investigate the critical temperature of a relativistic BEC driven by PRM, with bosons described by a complex scalar field. For non-interacting bosons, we find that the critical temperature can only be determined for a system with fixed angular momentum.  For such a system, the critical temperature is exactly zero for any value of the angular momentum, due to the fact that the system is \emph{quasi-one-dimensional} in PRM.  Therefore, \emph{non-interacting bosons cannot undergo Bose-Einstein condensation in PRM}. We then study a system of complex scalar field with quartic 
self-interaction in PRM. While previous numerical study shows that the mean-field ground state is a giant quantum vortex, here we demonstrate analytically that the bosons can in principle condense into several states with different angular quantum numbers; however, the true ground state of the BEC for realistic model parameters is still a giant vortex with a single angular quantum number. Then we take into account the phase fluctuation of the order parameter. We show that the order parameter actually vanishes and the off-diagonal long-range order is absent at any nonzero temperature because of the quasi-one-dimensional feature of the system, in accordance with the Coleman-Mermin-Wagner-Hohenberg theorem.

This paper is organized as follows. In Sec. II we set up a field theoretical model for relativistic bosons in PRM and evaluate the partition function in the imaginary-time formalism. In Sec. III we study the critical temperature for the non-interacting case. 
In Sec. IV we study the effect of phase fluctuation on the order parameter and the off-diagonal long-range order. We summarize in Sec. V.  The natural units $c=\hbar=k_{\rm B}=1$ are used throughout.

\section{The partition function at finite temperature}

Relativistic charged bosons can be described by a complex scalar field. The Lagrangian density is given by
\begin{equation}
\mathcal{L}=\left(\partial_\mu \Phi\right)^*\left(\partial^\mu\Phi\right) -m^2\Phi^*\Phi-V_{\rm int}(\Phi^*\Phi),
\end{equation}
with $m^2>0$ being the boson mass squared and $V$ the self-interaction. Assuming that the system is put in a constant magnetic field along the $z$ direction, $\boldsymbol{B}=B\hat {z}$.  Furthermore, 
a global rigid rotation along the magnetic field is applied, with angular velocity $\boldsymbol{ \Omega}=\Omega\hat{z}$.  
It is convenient to study the system in a rotating frame. The spacetime metric $g_{\mu\nu}$ of the rotating frame is given by
\begin{equation}
{\rm d}s^2=(1-\Omega^2\rho^2){\rm d}t^2+2\Omega y{\rm d}x{\rm d}t-2\Omega x {\rm d}y{\rm d}t-{\rm d}{\bf r}^2,
\end{equation}
where ${\bf r}=(x,y,z)$ and $\rho=\sqrt{x^2+y^2}$. The cylindrical coordinates ${\bf r}=(\rho,\theta,z)$ will also be used in the following. The action of the system is given by
\begin{equation}
{\cal S}=\int {\rm d}^4x\sqrt{-g} \Big[g^{\mu\nu}\left(D_\mu \Phi\right)^*\left(D_\nu\Phi\right) -m^2\Phi^*\Phi-V_{\rm int}(\Phi^*\Phi) \Big],
\end{equation}
where $g=\det(g_{\mu\nu})$. The magnetic field enters the Lagrangian density through the covariant derivative $D_\mu=\partial_\mu +i qA_\mu$, where $q$ is the charge and $A_\mu$ is the electromagnetic potential in the rotating frame. We consider the case $qB>0$ and $\Omega>0$ without loss of generality.  Since $\sqrt{-g}=1$, it is convenient to rewrite the action as ${\cal S}=\int {\rm d}^4x{\cal L}$, with the Lagrangian density 
\begin{equation}\label{Lag-R}
\mathcal{L}=|(D_t+\Omega y D_x - \Omega x D_y)\Phi|^2 - |D_i\Phi|^2 -m^2\Phi^*\Phi-V_{\rm int}(\Phi^*\Phi) .
\end{equation}

It is convenient to use the symmetric gauge so that the rotational symmetry along the $z$-axis is manifested. In the rest frame,  the electromagnetic potential is given by $A_\mu^{\rm R}=(0, By_{\rm R}/2,-Bx_{\rm R}/2,0)$, while in the rotating frame,  it becomes $A_\mu=(-B\Omega \rho^2/2, By/2,-Bx/2,0)$ according to the coordinate transformation $t_{\rm R}=t$, $\rho_{\rm R}=\rho$, $\theta_{\rm R}=\theta+\Omega t$. An additional electric field ${\bf E} = B\Omega \mbox{\boldmath{$\rho$}}$ is induced in the rotating frame. However, according to the identity
$D_t+\Omega y D_x - \Omega x D_y = \partial_t + \Omega y \partial_x - \Omega x \partial_y$, the induced electric field ${\bf E}$ cancels out automatically, indicating that the rotating frame corresponds only to a frame change with no new force~\cite{PRL120:032001}. Therefore, 
the Lagrangian density (\ref{Lag-R}) reduces to
\begin{eqnarray}
\mathcal{L} = |(\partial_t-i\Omega \hat{l}_z)\Phi|^2 - |D_i\Phi|^2 -m^2\Phi^*\Phi-V_{\rm int}(\Phi^*\Phi) ,
\end{eqnarray}
where  $\hat{l}_z \equiv -i(x\partial_y-y\partial_x) = -i \partial_\theta$ is the angular momentum operator along the $z$ direction.  Defining $\Phi=(\phi_1+i\phi_2)/\sqrt{2}$ as usual, we obtain
\begin{eqnarray}
\mathcal{L} &=& \frac{1}{2}\left\{\dot{\phi}_1^2+\dot{\phi}_2^2-2i\Omega(\dot{\phi}_1\hat{l}_z\phi_1+\dot{\phi}_2\hat{l}_z\phi_2)-\Omega^2\left[(\hat{l}_z\phi_1)^2+(\hat{l}_z\phi_2)^2\right]\right\}-\frac{m^2}{2}(\phi_1^2+\phi_2^2)\nonumber\\
&-&\frac{1}{2}\left[(\nabla\phi_1)^2+(\nabla\phi_2)^2+\frac{q^2B^2\rho^2}{4}(\phi_1^2+\phi_2^2)-iqB(\phi_1\hat{l}_z\phi_2-\phi_2\hat{l}_z\phi_1)\right]-V_{\rm int}\left(\frac{\phi_1^2+\phi_2^2}{2}\right).
\end{eqnarray}
Following the standard treatment, we define the conjugate fields
\begin{eqnarray}
\pi_1=\frac{\partial{\cal L}}{\partial\dot{\phi}_1}=\dot{\phi}_1-i\Omega\hat{l}_z\phi_1,\ \ \ \ \pi_2=\frac{\partial{\cal L}}{\partial\dot{\phi}_2}=\dot{\phi}_2-i\Omega\hat{l}_z\phi_2.
\end{eqnarray}
The Hamiltonian density ${\cal H}=\pi_1\dot{\phi}_1+\pi_2\dot{\phi}_2-{\cal L}$ can be evaluated as
\begin{eqnarray}
{\cal H}&=&\frac{1}{2}(\pi_1^2+\pi_2^2)+\frac{1}{2}\left[(\nabla\phi_1)^2+(\nabla\phi_2)^2\right]+\frac{m^2}{2}(\phi_1^2+\phi_2^2)\nonumber\\
&+&V_{\rm int}\left(\frac{\phi_1^2+\phi_2^2}{2}\right)+\frac{q^2B^2\rho^2}{8}(\phi_1^2+\phi_2^2)-\frac{iqB}{2}(\phi_1\hat{l}_z\phi_2-\phi_2\hat{l}_z\phi_1)\nonumber\\
&+&i\Omega(\pi_1\hat{l}_z\phi_1+\pi_2\hat{l}_z\phi_2)
\end{eqnarray}

At finite temperature,  the partition function of the system is given by
\begin{equation}
\mathcal{Z}=\text{Tr}{\rm e}^{-\beta(\hat{H}-\mu\hat{Q})},
\end{equation}
where $\beta=1/T$, $\hat{Q}$ is the conserved charge associated with the U$(1)$ symmetry,
\begin{equation}
\hat{Q}=\int{\rm d}{\bf r}(\pi_1\phi_2-\pi_2\phi_1),
\end{equation}
and $\mu$ is the corresponding chemical potential. We note that the Hamiltonian can be expressed as
\begin{equation}
\hat{H}=\hat{H}_0-\Omega\hat{L}_z,
\end{equation}
where $\hat{H}_0$ is the Hamiltonian at $\Omega=0$ and 
\begin{equation}
\hat{L}_z=-i\int{\rm d}{\bf r}(\pi_1\hat{l}_z\phi_1+\pi_2\hat{l}_z\phi_2)
\end{equation}
is the angular momentum of the field system along the $z$ direction.  Therefore, the partition function can also be expressed as
\begin{equation}
\mathcal{Z}=\text{Tr}{\rm e}^{-\beta(\hat{H}_0-\Omega\hat{L}_z-\mu\hat{Q})}.
\end{equation}
The statistic averages of the U$(1)$ charge $\hat{Q}$ and angular momentum $\hat{L}_z$ are given by
\begin{equation}
Q\equiv\langle\hat{Q}\rangle=\frac{1}{\beta}\frac{\partial\ln{\cal Z}}{\partial \mu},\ \ \ \ \ \ \ L_z\equiv\langle\hat{L}_z\rangle=\frac{1}{\beta}\frac{\partial\ln{\cal Z}}{\partial \Omega}.
\end{equation}

In the imaginary-time formalism, the partition function can be converted to a path integral
\begin{eqnarray}
{\cal Z}=\int{\cal D}\pi_1{\cal D}\pi_2\int{\cal D}\phi_1{\cal D}\phi_2\exp{\left[\int_X\left(i\pi_1\partial_\tau\phi_1+i\pi_2\partial_\tau\phi_2-{\cal H}+\mu{\cal Q}\right)\right]},
\end{eqnarray}
where $\int_X\equiv\int_0^\beta {\rm d}\tau\int {\rm d}{\bf r}$ and ${\cal Q}=\pi_1\phi_2-\pi_2\phi_1$. Completing the integrals over $\pi_1$ and $\pi_2$ yields
\begin{eqnarray}
{\cal Z}=\int{\cal D}\phi_1{\cal D}\phi_2\exp{\left(\int_X{\cal L}_{\rm eff}\right)},
\end{eqnarray}
where
\begin{eqnarray}
{\cal L}_{\rm eff}&=&\frac{1}{2}\left(i\partial_\tau\phi_1+\mu\phi_2-i\Omega\hat{l}_z\phi_1\right)^2+\frac{1}{2}\left(i\partial_\tau\phi_2-\mu\phi_1-i\Omega\hat{l}_z\phi_2\right)^2\nonumber\\
&-&\frac{1}{2}\left[(\nabla\phi_1)^2+(\nabla\phi_2)^2\right]-\frac{m^2}{2}(\phi_1^2+\phi_2^2)-V_{\rm int}\left(\frac{\phi_1^2+\phi_2^2}{2}\right)\nonumber\\
&-&\frac{q^2B^2\rho^2}{8}(\phi_1^2+\phi_2^2)+\frac{iqB}{2}(\phi_1\hat{l}_z\phi_2-\phi_2\hat{l}_z\phi_1).
\end{eqnarray}
Integrating by parts and using the complex field $\Phi$, we can express it as
\begin{eqnarray}
{\cal L}_{\rm eff}=\Phi^*\left[\left(\partial_{\tau}-\mu-\Omega \hat{l}_{z}\right)^{2}+\hat{K}_{\rm 2D}+\partial_{z}^{2}-m^{2}\right]\Phi-V(\Phi^*\Phi)
\end{eqnarray}
where the operator $K_{\rm 2D}$ is defined as:
\begin{equation}
\hat{K}_{\rm 2D}=\frac{\partial^{2}}{\partial\rho^{2}}+\frac{1}{\rho}\frac{\partial}{\partial\rho}-\frac{\hat{l}_{z}^{2}}{\rho^{2}}-\frac{1}{4}q^{2}B^{2}\rho^{2}+qB\hat{l}_{z}.\label{K_2D}
\end{equation}

\section{Non-Interacting Bosons in PRM}

In this section, we consider the non-interacting system, $V_{\rm int}=0$.  As a comparison, we first consider the case $B=\Omega=0$, i.e., non-interacting bosons in three-dimensional free space.  In this case, we have
\begin{eqnarray}
{\cal L}_{\rm eff}=\Phi^*\left[\left(\partial_{\tau}-\mu\right)^{2}+\nabla^{2}-m^{2}\right]\Phi,
\end{eqnarray}
leading to the equation of motion 
\begin{equation}
\left[\left(\partial_{\tau}-\mu\right)^{2}+\nabla^{2}-m^{2}\right]\Phi(\tau,{\bf r})=0.
\end{equation}
The eigen solution is simply given by the plane-wave form
\begin{equation}
\Phi(\tau,{\bf r})={\rm e}^{-\varepsilon\tau+i{\bf k}\cdot{\bf r}},\ \ \ \ \ (\varepsilon+\mu)^2={\bf k}^2+m^2.
\end{equation}
Therefore, to carry out the path integral, we expand the complex field as
\begin{equation} \label{ansatz_1}
\Phi(\tau,{\bf r})=\sqrt{\frac{\beta}{L^3}}\sum_{n}\sum_{\bf k}{\rm e}^{i\omega_n\tau+i{\bf k}\cdot{\bf r}}\tilde{\Phi}_{n{\bf k}},
\end{equation}
where $\omega_n=2n\pi T$ is the boson Matsubara frequency according to the periodic boundary condition $\Phi(0,{\bf r})=\Phi(\beta,{\bf r})$. We first put the system in a cubic box with volume $V=L^3$ and finally set $L\rightarrow\infty$.
Substituting this expansion into the action,  we obtain
\begin{eqnarray}\label{Action-free-space}
\int_X{\cal L}_{\rm eff}=-\sum_{n,{\bf k}}\beta^{2}\Big[(\omega_{n}-i\mu)^{2}+{\bf k}^2+m^2\Big]\tilde{\Phi}_{n{\bf k}}^{*}\tilde{\Phi}_{n{\bf k}}^{\phantom{\dag}}.
\end{eqnarray}
Finally, the variables of the path integral can be converted to be over $\tilde{\Phi}_{n{\bf k}}^{*}$ and $\tilde{\Phi}_{n{\bf k}}^{\phantom{\dag}}$.  In the absence of interaction, the integral is Gaussian and can be completed to obtain
\begin{equation}
\mathcal{Z}=\prod_{n,{\bf k}}\left[\frac{\omega_{n}^{2}+(E_{\bf k}-\mu)^{2}}{T^{2}}\right]^{-\frac{1}{2}} \left[\frac{\omega_{n}^{2}+(E_{\bf k}+\mu)^{2}}{T^{2}}\right]^{-\frac{1}{2}},
\end{equation}
where $E_{\bf k}\equiv\sqrt{{\bf k}^2+m^2}$. The grand potential $\Omega_{\rm G}=-T\ln \mathcal{Z}$ can be evaluated by performing the Matsubara sum.  We obtain
\begin{equation}
\frac{\Omega_{\rm G}}{V}=\int\frac{\mathrm{d}^3{\bf k}}{(2\pi)^3}\left[E_{\bf k}+T\ln\left(1-e^{-\beta(E_{\bf k}-\mu)}\right)+T\ln\left(1-e^{-\beta(E_{\bf k}+\mu)}\right)\right].
\end{equation}
The statistic average of the U$(1)$ charge can be evaluated as
\begin{equation}
\frac{Q}{V}=\int\frac{\mathrm{d}^3{\bf k}}{(2\pi)^3}\left[\frac{1}{{\rm e}^{\beta(E_{\bf k}-\mu)}-1}-\frac{1}{{\rm e}^{\beta(E_{\bf k}+\mu)}-1}\right].
\end{equation}

For non-interacting many-boson systems, the simple way to judge whether BEC exists is to study the conserved quantities of the system. Without loss of generality, we consider the case $\mu>0$. It is clear that when $\mu=m$, the Bose-Einstein distribution function with energy $E_{\bf k}-\mu$ develops a singularity at ${\bf k}=0$. However, the critical temperature $T_{\rm c}$ cannot be determined as a function of $\mu$, since $\mu$ is fixed to be $\mu=m$ in the BEC phase and hence no longer a thermodynamic variable.  The correct way to determine $T_{\rm c}$ in this non-interacting system is to impose the conserved quantities. The U$(1)$ charge is actually composed of two contributions,
\begin{equation}
\frac{Q}{V}=\frac{Q_{\rm c}}{V}+\int\frac{\mathrm{d}^3{\bf k}}{(2\pi)^3}\left[\frac{1}{{\rm e}^{\beta(E_{\bf k}-\mu)}-1}-\frac{1}{{\rm e}^{\beta(E_{\bf k}+\mu)}-1}\right],
\end{equation}  
where $Q_{\rm c}$ is from the condensed particles, and the other term is from thermally excited particles and antiparticles. At $T=T_{\rm c}$, we have $Q_{\rm c}=0$ and $\mu=m$. Thus, the critical temperature is determined by 
\begin{equation}\label{3DTC}
\frac{Q}{V}=\int\frac{\mathrm{d}^3{\bf k}}{(2\pi)^3}\left[\frac{1}{{\rm e}^{(E_{\bf k}-m)/T_{\rm c}}-1}-\frac{1}{{\rm e}^{(E_{\bf k}+m)/T_{\rm c}}-1}\right].
\end{equation}  
In the nonrelativistic limit, $Q$ reduces to the particle number, and we recover the well-known result in the textbook. We note that, the existence of a nonzero $T_{\rm c}$ is due to the fact that the integral (\ref{3DTC}) is safe in the infrared,
\begin{equation}
\frac{1}{{\rm e}^{(E_{\bf k}-m)/T_{\rm c}}-1}\sim \frac{2mT_{\rm c}}{{\bf k}^2},\ \ \ \ \ \ \ |{\bf k}|\rightarrow0.
\end{equation}

The appearance of a condensate can be understood from a field theoretical point of view.  When $\mu=m$, the zero mode term in the action (\ref{Action-free-space}) vanishes, that is, we have 
\begin{eqnarray}
(\omega_{n}-i\mu)^{2}+{\bf k}^2+m^2=0
\end{eqnarray}
for $n=0$ and ${\bf k}=0$. Therefore, the variables $\tilde{\Phi}_{00}^{\phantom{\dag}}$ and $\tilde{\Phi}^*_{00}$ become cyclical variables and should be separated when performing the path integral. They are actually proportional to the expectation value of the field operator $\hat{\Phi}$, i.e., $\langle\hat{\Phi}\rangle\equiv\zeta$, which characterizes the spontaneous breaking of the U$(1)$ symmetry.  Without loss of generality, we set $\zeta$ to be real and hence 
$\tilde{\Phi}_{00}^{\phantom{\dag}}=\zeta\sqrt{TV}$. After the separation of the zero mode contribution,  the partition function reads
\begin{equation}
\mathcal{Z}=\exp{\left[-\frac{V}{T}(m^2-\mu^2)\zeta^2\right]}\prod_{(n,{\bf k})\neq(0,0)}\left[\frac{\omega_{n}^{2}+(E_{\bf k}-\mu)^{2}}{T^{2}}\right]^{-\frac{1}{2}} \left[\frac{\omega_{n}^{2}+(E_{\bf k}+\mu)^{2}}{T^{2}}\right]^{-\frac{1}{2}}.
\end{equation}
The grand potential can be evaluated as
\begin{equation}
\frac{\Omega_{\rm G}}{V}=(m^2-\mu^2)\zeta^2+\int\frac{\mathrm{d}^3{\bf k}}{(2\pi)^3}\left[E_{\bf k}+T\ln\left(1-e^{-\beta(E_{\bf k}-\mu)}\right)+T\ln\left(1-e^{-\beta(E_{\bf k}+\mu)}\right)\right].
\end{equation}
The U$(1)$ charge becomes
\begin{equation}
\frac{Q}{V}=2\mu\zeta^2+\int\frac{\mathrm{d}^3{\bf k}}{(2\pi)^3}\left[\frac{1}{{\rm e}^{\beta(E_{\bf k}-\mu)}-1}-\frac{1}{{\rm e}^{\beta(E_{\bf k}+\mu)}-1}\right].
\end{equation}
It is clear that when $\mu=m$, the condensate $\zeta$ and the critical temperature $T_{\rm c}$ can only be determined by imposing the U$(1)$ charge $Q$.

Now we consider the system in PRM. To evaluate the partition function, we first solve the equation of motion
\begin{equation}\label{EMO}
\left[\left(\partial_{\tau}-\mu-\Omega \hat{l}_{z}\right)^{2}+\hat{K}_{\rm 2D}+\partial_{z}^{2}-m^{2}\right]\Phi(\tau,\rho,\theta,z)=0.
\end{equation}
The eigen solution can be written as
\begin{equation} \label{ansatz_2}
\Phi(\tau,\rho,\theta,z)={\rm e}^{-\varepsilon\tau+il\theta+ik_zz}\varphi(\rho).
\end{equation}
Substituting it into (\ref{EMO}), we obtain an eigenvalue equation for the function $\varphi(\rho)$,
\begin{equation}
\left[\frac{\partial^{2}}{\partial\rho^{2}}+\frac{1}{\rho}\frac{\partial}{\partial\rho}-\frac{l^{2}}{\rho^{2}}-\frac{1}{4}q^{2}B^{2}\rho^{2}\right]\varphi(\rho)=\left[k_z^2+m^2-qBl-(\varepsilon-\mu-l\Omega )^2\right]\varphi(\rho),
\end{equation}
To determine the radial eigenfunction $\varphi(\rho)\equiv\varphi_{sl}(\rho)$, an boundary condition is needed. Two kinds of boundary conditions are usually used: one is to impose the Dirichlet boundary condition at 
$\rho=R$, $\varphi_{sl}(R)=0$; the other is to require that the radial part $\varphi_{sl}(\rho)$ is square-integrable over the entire regime $0\leq\rho<\infty$. If we choose the first one, the radial solution takes the form
\begin{equation}
\varphi_{sl}(\rho)=\mathcal{N}_{sl}\rho^{|l|}{\rm e}^{-\frac{1}{4}qB\rho^2}M\left(a_{sl},|l|+1,\frac{qB\rho^2}{2}\right),\label{radius_EMO}
\end{equation}
where $\mathcal{N}_{nl}$ is a normalization factor and $M(a,b,x)$ is the Kummer Confluent Hypergeometric function with the parameter $a_{sl}$ determined by
\begin{equation}
M\left(a_{sl},|l|+1,\frac{qBR^2}{2}\right)=0.
\end{equation}
The Dirichlet boundary condition is convenient to impose an exact speed of light constraint $\Omega R<1$.  However,  the Kummer function $M(a,b,x)$ is too complicated for further analytical and numerical calculations.  The asymptotic behavior of the Kummer function $M(a,b,x)$ at $x\to\infty$ is \cite{olver_nist_2010}
\begin{equation}
M(a,b,x)\sim\frac{{\rm e}^{x}x^{a-b}}{{\Gamma(a)}}\sum_{s=0}^{\infty}\frac{(1-a)_{s}(b-a)_{s}}{s!}x^{-s}.
\end{equation}
Thus, the radial function $\varphi_{sl}(\rho)$ does not vanish at $\rho\to\infty$. Near the boundary, the radial function $\varphi_{sl}(\rho)$ oscillates intensively, which is not convenient for numerical calculations.

If we take another boundary condition, i.e., the radial function $\varphi_{sl}(\rho)$ is square-integrable over the entire regime $0\leq\rho<\infty$, the solution takes the form 
\begin{equation}
\varphi_{sl}(\rho)=\Bigl(\frac{1}{2}qB\rho^{2}\Bigr)^{|l|/2}{\rm e}^{-\frac{1}{4}qB\rho^{2}}L_{s}^{(|l|)}\Bigl(\frac{1}{2}qB\rho^{2}\Bigr),\label{modified_wave_function}
\end{equation}
where $L_{s}^{(|l|)}$ are the Laguerre polynomials ($s=0,1,2,\cdots$). The corresponding eigenvalues (Landau levels) take a very simple analytical form
\begin{equation}
(\varepsilon_{sl}-\mu-l\Omega)^{2}=qB(2s+|l|-l+1)+k_{z}^{2}+m^{2}.
\end{equation}
The Laguerre polynomials and the analytical eigenvalues are very convenient for further analytical calculations. The cost is that the speed of light constraint $\Omega R<1$ cannot be imposed exactly.  However, since the radial function $\varphi_{sl}(\rho)$ vanishes fast at $\rho\to\infty$,  this problem may be less important.  For a cylindrical system with radius $R$, the degeneracy of the Landau level is given by 
\begin{equation}
N=\frac{qBS}{2\pi}=\frac{1}{2}qBR^2,
\end{equation}
with $S=\pi R^{2}$ being the area of the system perpendicular to the external magnetic field. In this work, we assume that the magnetic field is sufficiently strong, so that the magnetic length $1/\sqrt{qB}\ll R$ for the Landau levels to fit within the area $S$. Thus, the azimuthal quantum number $l$ is constrained in the range $-s\leqslant l \leqslant N-s$, where the degeneracy $N\gg 1$.

Then we calculate the Gaussian path integral in the partition function ${\cal Z}$ by using the orthogonal basis functions $\varphi_{sl}(\rho)$. We expand the complex field $\Phi(\tau,\rho,\theta,z)$ as
\begin{equation}\label{Eq44}
\Phi(\tau,\rho,\theta,z)=\sqrt{\frac{qB\beta}{2\pi H_{z}}}\sum_{K}\tilde{\Phi}_K\sqrt{\frac{s!}{(s+|l|)!}}{\rm e}^{i\omega_{n}\tau+il\theta+ik_{z}z}\varphi_{sl}(\rho),       
\end{equation}
where $H_z$ is the length of system along the $z$ direction and $K$ denotes the set of quantum numbers $\{n,s,l,k_z\}$. The summation over $K$ is explicitly defined as
\begin{equation}
\sum_K\equiv H_{z}\sum_{n=-\infty}^{n=\infty}\sum_{s=0}^{\infty}\sum_{l=-s}^{N-s}\int_{-\infty}^{\infty}\frac{\mathrm{d}k_{z}}{2\pi}.
\end{equation}
Substituting this expansion into the action and $K^{*}$ denotes $\{-n,s,-l,-k_{z}\}$, we obtain
\begin{eqnarray}
\int_X{\cal L}_{\rm eff}=-\sum_{K}\beta^{2}\Big[(\omega_{n}-i\mu-il\Omega)^{2}+qB(2s+|l|-l+1)+k_z^2+m^2\Big]\tilde{\Phi}_{K^{*}}\tilde{\Phi}_{K}^{\phantom{\dag}}
\end{eqnarray}
Thus, the thermal propagator of the complex field in the $K$-space is diagonal and can be given by
\begin{equation}
{\cal D}_0(K)=\frac{1}{(\omega_{n}-i\mu-il\Omega)^{2}+qB(2s+|l|-l+1)+k_z^2+m^2}.
\end{equation}
Finally, the variables of the path integral can be converted to be over $\tilde{\Phi}_K^{*}$ and $\tilde{\Phi}_{K}^{\phantom{\dag}}$.  In the absence of interaction, the integral is Gaussian and can be completed to obtain
\begin{equation}
\mathcal{Z}=\prod_K\left[\frac{\omega_{n}^{2}+(E_{sl}-\mu-l\Omega)^{2}}{T^{2}}\right]^{-\frac{1}{2}} \left[\frac{\omega_{n}^{2}+(E_{sl}+\mu+l\Omega)^{2}}{T^{2}}\right]^{-\frac{1}{2}},
\end{equation}
where $E_{sl}$ is Landau energy
\begin{equation}
E_{sl}\equiv \sqrt{qB(2s+|l|-l+1)+k_{z}^{2}+m^2}.
\end{equation}
The grand potential $\Omega_{\rm G}=-T\ln \mathcal{Z}$ can be evaluated by performing the Matsubara sum.  We obtain
\begin{equation}\label{modified_free_part_thermal_potential}
\Omega_{\rm G}=H_{z}\sum_{s=0}^{\infty}\sum_{l=-s}^{N-s}\int_{-\infty}^{\infty}\frac{\mathrm{d}k_{z}}{2\pi}\left[E_{sl}+T\ln\left(1-e^{-\beta(E_{sl}-\mu-l\Omega)}\right)+T\ln\left(1-e^{-\beta(E_{sl}+\mu+l\Omega)}\right)\right].
\end{equation}
The statistic average of the U$(1)$ charge can be evaluated as
\begin{equation}\label{expressionQ_1}
Q=H_z\sum_{s=0}^{\infty}\sum_{l=-s}^{N-s}\int_{-\infty}^{\infty}\frac{\mathrm{d}k_{z}}{2\pi}\left[\frac{1}{{\rm e}^{\beta(E_{sl}-\mu-l\Omega)}-1}-\frac{1}{{\rm e}^{\beta(E_{sl}+\mu+l\Omega)}-1}\right].
\end{equation}
The statistic average of the angular momentum is given by
\begin{equation}\label{expressionLz}
L_{z}=H_z\sum_{s=0}^{\infty}\sum_{l=-s}^{N-s}\int_{-\infty}^{\infty}\frac{\mathrm{d}k_{z}}{2\pi}\left[\frac{l}{{\rm e}^{\beta(E_{sl}-\mu-l\Omega)}-1}-\frac{l}{{\rm e}^{\beta(E_{sl}+\mu+l\Omega)}-1}\right].
\end{equation}

%%%%%%%%%%%%%%%%%%%%%%%%%%%%%%%%%%%%%%%%%%%%%%%%%%%%%%%%%%%%%%%%%%%%%%%
\begin{figure}[!htb]
\begin{center}
\includegraphics[width=0.4\linewidth]{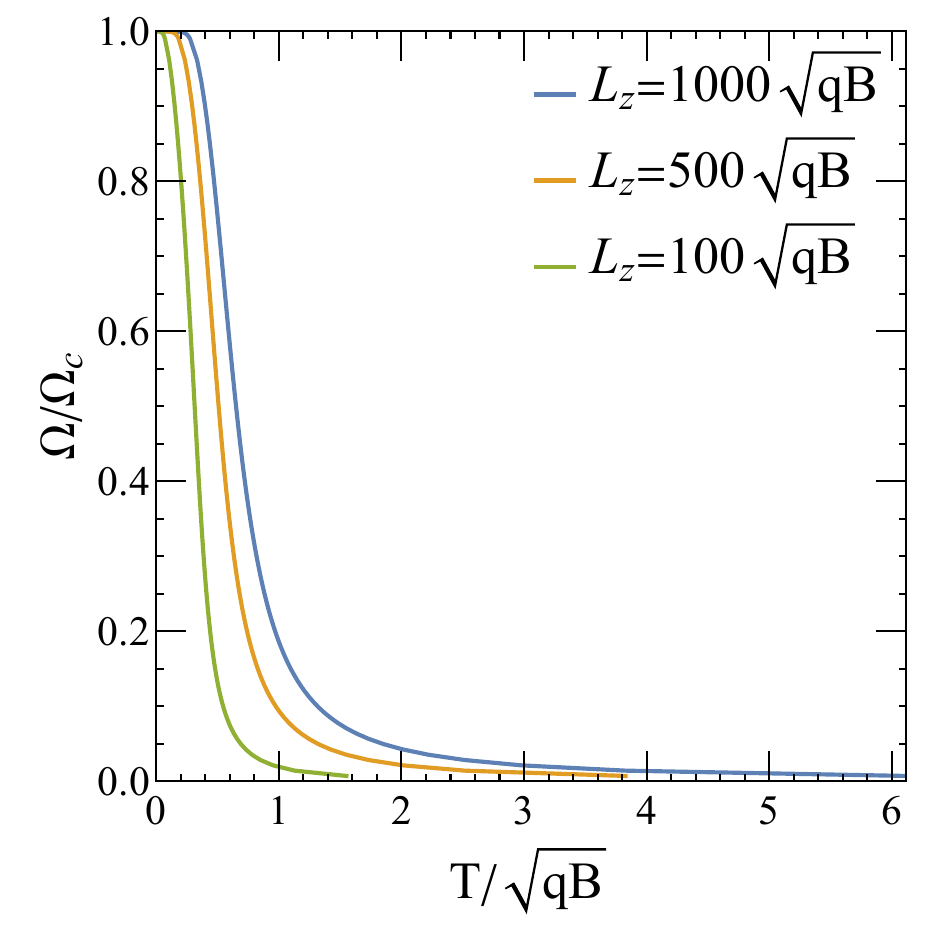}
\caption{\label{fig01}The angular speed $\Omega$ as a function of $T$ for different values of the angular momentum $L_{z}$ at $\mu=0$. In the numerical calculation, we set $m^{2}=qB$ and $N=100$.}
\end{center}
\end{figure}
%%%%%%%%%%%%%%%%%%%%%%%%%%%%%%%%%%%%%%%%%%%%%%%%%%%%%%%%%%%%%%%%%%%%%%%%

The present system in PRM has two conserved quantities, the $U(1)$ charge $Q$ and the angular momentum $L_z$.  Their expressions (\ref{expressionQ_1}) and (\ref{expressionLz}) are sums (integrals) of the Bose-Einstein distribution functions
\begin{equation}
f_{\rm B}^-=\frac{1}{{\rm e}^{\beta(E_{sl}-\mu-l\Omega)}-1},\ \ \ \ \ \ \ \ \ \ f_{\rm B}^+=\frac{1}{{\rm e}^{\beta(E_{sl}+\mu+l\Omega)}-1}.
\end{equation} 
A singularity appears if $E_{sl}-\mu-l\Omega=0$ or $E_{sl}+\mu+l\Omega=0$. For the sake of simplicity, we first consider the case $\mu=0$. In this case, we do not need to impose the U$(1)$ 
charge $Q$. The Landau energy $E_{sl}$ has a minimum at $k_z=0,s=0$ and $l\geqslant0$, i.e.,
\begin{equation}
\min E_{sl}=\sqrt{qB+m^2}.
\end{equation} 
For the lowest Landau level $s=0$, the azimuthal quantum number $l$ is constrained in the range $0\leqslant l \leqslant N$.  The BEC is expected to occur when $N\Omega=\sqrt{qB+m^2}$, i.e., particles with azimuthal quantum number $l=N$ in the lowest Landau level become condensed.  However, as in the free space case, we cannot determine the 
critical temperature $T_{\rm c}$ as a function of $\Omega$, since its value is fixed to be
\begin{equation}\label{Eq55}
\Omega=\Omega_{\rm c}=\frac{\sqrt{qB+m^2}}{N}.
\end{equation} 
The correct way determine $T_{\rm c}$ is to impose the angular momentum $L_z$, which is composed from two contributions,
\begin{equation}
L_{z}=L_{z\rm c}+H_z\sum_{s=0}^{\infty}\sum_{l=-s}^{N-s}\int_{-\infty}^{\infty}\frac{\mathrm{d}k_{z}}{2\pi}\left[\frac{l}{{\rm e}^{\beta(E_{sl}-l\Omega)}-1}-\frac{l}{{\rm e}^{\beta(E_{sl}+l\Omega)}-1}\right].
\end{equation}
Here $L_{z\rm c}$ denotes the angular momentum from the condensed particles. At $T=T_{\rm c}$, we have $L_{z\rm c}=0$ and $\Omega=\Omega_{\rm c}$. Thus, the critical temperature $T_{\rm c}$ is determined by 
\begin{equation}
L_{z}=H_z\sum_{s=0}^{\infty}\sum_{l=-s}^{N-s}\int_{-\infty}^{\infty}\frac{\mathrm{d}k_{z}}{2\pi}\left[\frac{l}{{\rm e}^{(E_{sl}-l\Omega_{\rm c})/T_{\rm c}}-1}-\frac{l}{{\rm e}^{(E_{sl}+l\Omega_{\rm c})/T_{\rm c}}-1}\right].
\end{equation}
However, this equation gives a definite result
\begin{equation}
T_{\rm c}=0.
\end{equation}
To prove this result, we note that for $s=0$ and $l=N$, we have
\begin{equation}
E_{0N}-N\Omega_{\rm c}=\sqrt{k_z^2+qB+m^2}-\sqrt{qB+m^2}=\frac{k_z^2}{2\sqrt{qB+m^2}}+O(k_z^4)
\end{equation}
Using the expansion for the Bose-Einstein distribution
\begin{equation}
\frac{1}{{\rm e}^{E/T}-1}=\frac{T}{E}-\frac{1}{2}+O(E),\ \ \ \ \ E\rightarrow 0,
\end{equation}
we find that the infrared behavior of the integral over $k_z$ is
\begin{equation}
\int_{-\delta}^{\delta}\frac{\mathrm{d}k_{z}}{2\pi}\frac{1}{{\rm e}^{(E_{0N}-N\Omega_{\rm c})/T_{\rm c}}-1}\sim 2T\sqrt{qB+m^2}\int_{-\delta}^{\delta}\frac{\mathrm{d}k_{z}}{2\pi}\frac{1}{k_z^2}.
\end{equation}
It is infrared divergent. Therefore, $T_{\rm c}$ must vanish.  In Fig. \ref{fig01}, we calculate the angular speed $\Omega$ as a function of the temperature $T$ for different values of $L_z$.  
The numerical results indicate that for arbitrary values of $L_z$, angular speed $\Omega$ approaches its critical value $\Omega_{\rm c}$ at $T=0$.  Thus, the system does not Bose condense at any nonzero temperature $T\neq0$.
The physical picture now becomes clear: In a magnetic field, the motion of the particles becomes effectively one-dimensional.

For general cases, we should also impose the U$(1)$ charge $Q$.  We have
\begin{eqnarray}
&&Q=Q_{\rm c}+H_z\sum_{s=0}^{\infty}\sum_{l=-s}^{N-s}\int_{-\infty}^{\infty}\frac{\mathrm{d}k_{z}}{2\pi}\left[\frac{1}{{\rm e}^{\beta(E_{sl}-\mu-l\Omega)}-1}-\frac{1}{{\rm e}^{\beta(E_{sl}+\mu+l\Omega)}-1}\right],\nonumber\\
&&L_{z}=L_{z\rm c}+H_z\sum_{s=0}^{\infty}\sum_{l=-s}^{N-s}\int_{-\infty}^{\infty}\frac{\mathrm{d}k_{z}}{2\pi}\left[\frac{l}{{\rm e}^{\beta(E_{sl}-\mu-l\Omega)}-1}-\frac{l}{{\rm e}^{\beta(E_{sl}+\mu+l\Omega)}-1}\right].
\end{eqnarray}
Assuming $\mu+N\Omega>0$, the BEC is expected to occur when
\begin{eqnarray}\label{Mucase01}
\mu_{\rm c}+N\Omega_{\rm c}=\sqrt{qB+m^2}.
\end{eqnarray}
As in the $\mu=0$ case, particles with azimuthal quantum number $l=N$ in the lowest Landau level $s=0$ become condensed. At $T=T_{\rm c}$, we have $Q_{\rm c}=0$ and $L_{z\rm c}=0$, leading to 
\begin{eqnarray}\label{Mucase02}
&&Q=H_z\sum_{s=0}^{\infty}\sum_{l=-s}^{N-s}\int_{-\infty}^{\infty}\frac{\mathrm{d}k_{z}}{2\pi}\left[\frac{1}{{\rm e}^{(E_{sl}-\mu_{\rm c}-l\Omega_{\rm c})/T_{\rm c}}-1}-\frac{1}{{\rm e}^{(E_{sl}+\mu_{\rm c}+l\Omega_{\rm c})/T_{\rm c}}-1}\right],
\nonumber\\
&&L_{z}=H_z\sum_{s=0}^{\infty}\sum_{l=-s}^{N-s}\int_{-\infty}^{\infty}\frac{\mathrm{d}k_{z}}{2\pi}\left[\frac{l}{{\rm e}^{(E_{sl}-\mu_{\rm c}-l\Omega_{\rm c})/T_c}-1}-\frac{l}{{\rm e}^{(E_{sl}+\mu_{\rm c}+l\Omega_{\rm c})/T_c}-1}\right].
\end{eqnarray}
The critical temperature $T_{\rm c}$ and the values of $\mu_{\rm c}$ and $\Omega_{\rm c}$ can be determined by Eqs. (\ref{Mucase01}) and (\ref{Mucase02}).  Because the integrals over $k_z$ are also infrared divergent for any $T\neq0$, we also have $T_{\rm c}$=0. In realistic systems we may require charge neutrality $Q=0$.  At sufficiently low temperature, $T\ll\sqrt{qB}$, we can keep only the lowest Landau level to write
\begin{equation}\label{expressionQ_2}
\sum_{l=0}^{N}\int_{-\infty}^{\infty}\frac{\mathrm{d}k_{z}}{2\pi}\left[\frac{1}{{\rm e}^{\beta(E_{0l}-\mu-l\Omega)}-1}-\frac{1}{{\rm e}^{\beta(E_{0l}+\mu+l\Omega)}-1}\right]=0.
\end{equation}
This equation has a simple analytical solution $\mu=-N\Omega/2$ \cite{PRL120:032001}, which validates our assumption $\mu+N\Omega>0$. However, at high temperature, the chemical potential $\mu$ deviates significantly from this result because of the 
thermal excitations of the higher Landau levels.

The mechanism of a possible BEC in PRM can also be understood from a field theoretical point of view.  We write down again the action
\begin{eqnarray}
\int_X{\cal L}_{\rm eff}=-\sum_{n,s,l,k_z}\beta^{2}\Big[(\omega_{n}-i\mu-il\Omega)^{2}+qB(2s+|l|-l+1)+k_z^2+m^2\Big]\tilde{\Phi}_{n,s,l,k_z}^{*}\tilde{\Phi}_{n,s,l,k_z}^{\phantom{\dag}}
\end{eqnarray}
For the mode $(n,s,l,k_z)=(0,0,N,0)$, the action is
\begin{eqnarray}
{\cal S}_{0,0,N,0}=-\beta^{2}\Big[qB+m^2-(\mu+N\Omega)^{2}\Big]\tilde{\Phi}_{0,0,N,0}^{*}\tilde{\Phi}_{0,0,N,0}^{\phantom{\dag}}
\end{eqnarray}
Therefore, if $\mu+N\Omega=\sqrt{qB+m^2}$, the above action vanishes and the variables $\tilde{\Phi}_{0,0,N,0}^{\phantom{\dag}}$ and $\tilde{\Phi}^*_{0,0,N,0}$ become cyclical variables.  We write
\begin{eqnarray}
\tilde{\Phi}_{0,0,N,0}=v\sqrt{\frac{2\pi H_zT}{qB}}
\end{eqnarray}
and set $\zeta$ to be real without loss of generality, which is equivalent to setting the expectation value of the field operator to be
\begin{eqnarray}
\langle\hat{\Phi}\rangle=v\sqrt{\frac{1}{N!}}{\rm e}^{iN\theta}\varphi_{0N}(\rho).
\end{eqnarray}
After separating the $(0,0,N,0)$ mode, we can evaluate the partition function as
\begin{eqnarray}
 \mathcal{Z}&=&\exp\left\{-\frac{2\pi H_{z}}{qBT}\left[qB+m^{2}-(\mu+N\Omega)^{2}\right]v^{2}\right\}\nonumber\\
 &&\times\ \prod_{(n,s,l,k_{z})\neq(0,0,N,0)}\left[\frac{\omega_{n}^{2}+(E_{sl}-\mu-l\Omega)^{2}}{T^{2}}\right]^{-\frac{1}{2}} \left[\frac{\omega_{n}^{2}+(E_{sl}+\mu+l\Omega)^{2}}{T^{2}}\right]^{-\frac{1}{2}}.
\end{eqnarray}
The grand potential $\Omega_{\rm G}=-T\ln{\cal Z}$ is calculated by completing the Matsubara sum. In the limit $H_z\rightarrow\infty$, we can add the $(0,0, N,0)$ mode to the sum again. The grand potential is given by
\begin{eqnarray}
\Omega_{\rm G}&=&H_z\sum_{s=0}^{\infty}\sum_{l=-s}^{N-s}\int_{-\infty}^{\infty}\frac{\mathrm{d}k_{z}}{2\pi}\left[E_{sl}+T\ln\left(1-{\rm e}^{-\beta(E_{sl}-\mu-l\Omega)}\right)+T\ln\left(1-{\rm e}^{-\beta(E_{sl}+\mu+l\Omega)}\right)\right]\nonumber\\
 &&+\ \frac{2\pi H_z}{qB}\left[qB+m^{2}-(\mu+N\Omega)^{2}\right]v^{2}
\end{eqnarray}
The U$(1)$ charge and the angular momentum can be evaluated as
\begin{eqnarray}
&&Q=\frac{4\pi H_z}{qB}(\mu+N\Omega)v^2+H_z\sum_{s=0}^{\infty}\sum_{l=-s}^{N-s}\int_{-\infty}^{\infty}\frac{\mathrm{d}k_{z}}{2\pi}\left[\frac{1}{{\rm e}^{\beta(E_{sl}-\mu-l\Omega)}-1}-\frac{1}{{\rm e}^{\beta(E_{sl}+\mu+l\Omega)}-1}\right],\nonumber\\
&&L_{z}=\frac{4\pi H_z}{qB}(\mu+N\Omega)Nv^2+H_z\sum_{s=0}^{\infty}\sum_{l=-s}^{N-s}\int_{-\infty}^{\infty}\frac{\mathrm{d}k_{z}}{2\pi}\left[\frac{l}{{\rm e}^{\beta(E_{sl}-\mu-l\Omega)}-1}-\frac{l}{{\rm e}^{\beta(E_{sl}+\mu+l\Omega)}-1}\right].
\end{eqnarray}
Therefore, we have $L_{z\rm c}=NQ_{\rm c}$. At $T=T_{\rm c}$, $v=0$. Thus, we obtain the same conclusion $T_{\rm c}=0$.

In the above calculations, we have used a convenient boundary condition that makes the results analytical. Obviously, other boundary conditions (such as the Dirichlet boundary condition) do not lead to a qualitatively different conclusion.  The reason for the absence of BEC is the quasi-one-dimensional feature of the system, which is independent of the boundary condition.

\section{Interacting Bosons in PRM}

Now we turn on the interaction. We consider a quartic  self-interaction
\begin{equation}
V_{\rm int}=\lambda(\Phi^*\Phi)^2.
\end{equation}
The most general ansatz for the classical part of the field $\Phi(X)$ is
\begin{equation}
\Phi_{\rm c}=\sum_{l=0}^Nv_l\sqrt{\frac{1}{|l|!}}{\rm e}^{il\theta}\varphi_{0l}(\rho)\label{classical_solution}
\end{equation}
Substituting it into the action, we obtain the effective potential at the tree-level 
\begin{equation}
\mathcal{V}_{\text{eff}}^{(0)}(v_l)=\frac{qBT}{2\pi H_{z}}\int_X{\cal L}(\Phi\rightarrow\Phi_{\rm c})
\end{equation}
It can be evaluated as
\begin{equation}
\mathcal{V}_{\text{eff}}^{(0)}(v_l)=\sum_{l=0}^N\left[qB+m^{2}-(\mu+l\Omega)^2\right]v_{l}^{2}+\lambda\sum_{l_1l_2l_3l_4}C_{l_1l_2l_3l_4}v_{l_1}v_{l_2}v_{l_3}v_{l_4}.\label{effective_potential}
\end{equation}
Using $L_{0}^{(\alpha)}=1$, we have
\begin{equation}
    \begin{aligned}
        C_{l_1l_2l_3l_4}&=\delta_{l_{1}+l_{2},l_{3}+l_{4}}\sqrt{\frac{1}{|l_{1}|!}\frac{1}{|l_{2}|!}\frac{1}{|l_{3}|!}\frac{1}{|l_{4}|!}}
\times\int_{0}^{+\infty}\mathrm{d}xe^{-2x}x^{\frac{|l_{1}|+|l_{2}|+|l_{3}|+|l_{4}|}{2}}\\
&=\delta_{l_{1}+l_{2},l_{3}+l_{4}}2^{-\frac{|l_{1}|+|l_{2}|+|l_{3}|+|l_{4}|}{2}-1}\sqrt{\frac{1}{|l_{1}|!}\frac{1}{|l_{2}|!}\frac{1}{|l_{3}|!}\frac{1}{|l_{4}|!}}\Gamma\Bigl(\frac{|l_{1}|+|l_{2}|+|l_{3}|+|l_{4}|}{2}+1\Bigr).
    \end{aligned}
\end{equation}
We are interested in the case of a strong magnetic field, $qB\sim m^2$. Variational calculation shows that the minimum of the tree-level effective potential is located at the point where $v_N\neq0$ and all other $v_l$s are vanishingly small (see Appendix A). Therefore, it is safe to set $\Phi_{\rm c}$ to be
\begin{equation}
\Phi_{\rm c}=v\sqrt{\frac{1}{N!}}{\rm e}^{iN\theta}\varphi_{0N}(\rho),\label{ground_state}
\end{equation}
where $v\equiv v_N$. The tree-level effective potential becomes
\begin{equation}
\mathcal{V}_{\text{eff}}^{(0)}(v)=\left[qB+m^{2}-(\mu+N\Omega)^2\right]v^{2}+\lambda C_{NNNN}v^4
\end{equation}
The critical angular speed is determined by the quadratic term. For $\mu=0$, it is given by (\ref{Eq55}).

Now we consider the fluctuations. To this end, we express the complex field $\Phi(X)$ as
\begin{equation}\label{phase and amplitude fluctuation}
\Phi(X)=\Biggl[v\sqrt{\frac{1}{N!}}\varphi_{0N}(\rho)\Bigl(1+{\rm A}(X)\Bigr)\Biggr]{\rm e}^{iN\theta+i{\rm P}(X)},
\end{equation}
where ${\rm P}(X)$ and ${\rm A}(X)$ correspond to the phase and amplitude fluctuations, respectively. The Lagrangian ${\cal L}_{\rm eff}$ can be expressed in terms of  ${\rm P}(X)$ and ${\rm A}(X)$.  We are interested in how the gapless fluctuations influence the order parameter and the off-diagonal long-range order, and therefore focus on the phase fluctuations. The terms relevant to ${\rm P}(X)$ are given by
\begin{equation}
    \begin{aligned}
        \mathcal{L}_{\rm P}=& \ i\Phi_{\rm c}^*\Phi_{\rm c}^{\phantom{\dag}}\left[(\partial_{\tau}-\mu+i\Omega\partial_{\theta})^{2}-\mu^{2}+\partial_{\rho}^{2}+\frac{1}{\rho}\partial_{\rho}+\frac{1}{\rho^{2}}\partial_{\theta}^{2}-iqB\partial_{\theta}
        +\partial_{z}^{2}\right]{\rm P}(X)\\
        &+2i\Phi^{*}_{\rm c}(\partial_{\rho}\Phi_{\rm c}^{\phantom{\dag}})\partial_{\rho}{\rm P}(X)-2i\Phi^{*}_{\rm c}(\partial_{\theta}\Phi_{\rm c}^{\phantom{\dag}})\left(-\frac{1}{\rho^{2}}+\Omega^{2}\right)\partial_{\theta}{\rm P}(X)-2iv\Omega
        \Phi^{*}_{\rm c}(\partial_{\theta}\Phi_{\rm c}^{\phantom{\dag}})\partial_{t}{\rm P}(X)\\
        &-\Phi^{*}_{\rm c}\Phi_{\rm c}^{\phantom{\dag}}\left[(\partial_{\tau}{\rm P})^{2}+(\partial_{\rho}{\rm P})^{2}+2i\Omega(\partial_{\tau}{\rm P})(\partial_{\theta}{\rm P})+\left(\frac{1}{\rho^{2}}-\Omega^{2}\right)(\partial_{\theta}{\rm P})^{2}
        +(\partial_{z}{\rm P})^{2}\right].
    \end{aligned}
\end{equation}
The equation of motion for ${\rm P}(x)$ can be derived by making a variation on ${\rm P}(x)$ and using the fact that $\Phi^{*}_{\rm c}\Phi_{\rm c}^{\phantom{\dag}}$ is only a function of $\rho$. The result is
\begin{equation}\label{EOM of phase fluctuation_1}
\left[\partial_{\tau}^{2}+\partial_{\rho}^{2}+(\Phi_{\rm c}^{*}\Phi_{\rm c}^{\phantom{\dag}})^{-1}\partial_{\rho}(\Phi_{\rm c}^{*}\Phi_{\rm c}^{\phantom{\dag}})\partial_{\rho}+2i\Omega\partial_{\theta}\partial_{\tau}+\left(\frac{1}{\rho^{2}}-\Omega^{2}\right)\partial_{\theta}^{2}+\partial_{z}^{2}\right]{\rm P}(X)=0.
\end{equation}
The linear terms in ${\rm P}(X)$ are proportional to the left-hand side of this equation of motion up to some total derivatives, as we expected. Using the explicit form of $\Phi_{\rm c}$ and $\hat{l}_{z}=-i\partial_{\theta}$, we obtain
\begin{equation}\label{EOM of phase fluctuation_2}
\left[(\partial_{\tau}-\Omega\hat{l}_{z})^{2}+\partial_{\rho}^{2}+\left(\frac{2N}{\rho}-qB\rho\right)\partial_{\rho}+\frac{1}{\rho^{2}}\partial_{\theta}^{2}+\partial_{z}^{2}\right]{\rm P}(X)=0.
\end{equation}
The eigen solution of ${\rm P}(X)$ can be written as
\begin{equation}
    {\rm P}(\tau,\rho,\theta,z)=e^{-\varepsilon \tau+il\theta+ik_{z}z}\xi(\rho),
\end{equation}
and the eigenvalue equation for $\xi(\rho)$ reads
\begin{equation}\label{EOM of phase fluctuation_3}
    \left[\partial_{\rho}^{2}+\left(\frac{2N}{\rho}-qB\rho\right)\partial_{\rho}-\frac{l^{2}}{\rho^{2}}\right]\xi(\rho)=[k_{z}^{2}-(\varepsilon+\Omega l)^{2}]\xi(\rho).
\end{equation}
We impose the same boundary condition as for $\varphi(\rho)$, i.e., $\xi(\rho)$ is square-integrable over the entire regime $0<\rho<\infty$. The solution takes the form 
\begin{equation}
    \xi_{sl}(\rho)=\left(\frac{1}{2}qB\rho^{2}\right)^{\frac{1}{4}\left[1-2N+\sqrt{4l^{2}+(2N-1)^{2}}\right]}L_{s}^{\left(1+\frac{1}{2}\sqrt{4l^{2}+(2N-1)^{2}}\right)}\left(\frac{1}{2}qB\rho^{2}\right).
\end{equation}
The corresponding eigenvalues are given by
\begin{equation}\label{goldstone dispersion relation}
    (\varepsilon_{sl}+\Omega l)^{2}=\frac{1}{2}qB\left(1-2N+\sqrt{4l^{2}+(2N-1)^{2}}+4s\right)+k_{z}^{2}.
\end{equation}
Taking $s=l=0$,  we obtain $\varepsilon^{2}=k_{z}^{2}$,  i.e., the gapless Goldstone mode.

Now we expand the phase fluctuation ${\rm P}(X)$ in terms of the eigenfunctions, 
\begin{equation}
    {\rm P}(X)=\sqrt{\frac{qB\beta}{2\pi v^{2}H_{z}}}\sum_{K}\tilde{\rm P}_{K}e^{i\omega_{n}\tau+il\theta+ik_{z}z}\xi_{sl}(\rho).
\end{equation}
The action for ${\rm P}(X)$ can be computed to be
\begin{equation}
  S_{\rm P}=\int_X{\cal L}_{\rm P}=-\sum_{K}\beta^{2}\mathcal{G}^{-1}(K)\tilde{\rm P}_{K}^*\tilde{\rm P}_{K}^{\phantom{\dag}},
\end{equation}
where $\mathcal{G}(K)$ is the thermal propagator of the phase fluctuation,
\begin{equation}
    \mathcal{G}(K)=\frac{1}{(\omega_{n}+il\Omega)^{2}+\frac{1}{2}qB\left(1-2N+\sqrt{4l^{2}+(2N-1)^{2}}+4s\right)+k_{z}^{2}}.
\end{equation}
The expectation value of the phase factor ${\rm e}^{i{\rm P}(X)}$ is given by 
\begin{equation}
\langle{\rm e}^{i{\rm P}(X)}\rangle=\frac{\int{\cal D}{\rm P}{\rm e}^{i{\rm P}(X)}{\rm e}^{S_{\rm P}}}{\int{\cal D}{\rm P}{\rm e}^{S_{\rm P}}}.
\end{equation}
Since $S_{\rm P}$ is Gaussian, we obtain
\begin{equation}
\langle{\rm e}^{i{\rm P}(X)}\rangle={\rm e}^{-\frac{1}{2}\langle {\rm P}^{2}(X)\rangle}.
\end{equation}
On the other hand, the thermal propagator in the coordinate space is defined as
\begin{equation}
G(X,Y)=\langle {\rm P}(X){\rm P}(Y)\rangle=\frac{\int{\cal D}{\rm P}\left[{\rm P}(X){\rm P}(Y)\right]{\rm e}^{S_{\rm P}}}{\int{\cal D}{\rm P}{\rm e}^{S_{\rm P}}}.
\end{equation}
Therefore, we have
\begin{equation}
\langle {\rm P}^{2}(X)\rangle=G(X,X)=\frac{qBT}{2\pi v^{2}H_{z}}\sum_K\xi_{sl}^{2}(\rho){\cal G}(K).
\end{equation}
The contribution from the gapless Goldstone mode with $s=l=0$ can be evaluated as
\begin{equation}
 \frac{qBT}{2\pi v^{2}}\sum_{n=-\infty}^{\infty}\int_{-\infty}^{\infty}\frac{\mathrm{d}k_{z}}{2\pi}\frac{1}{\omega_{n}^{2}+k_{z}^{2}}=\frac{qB}{4\pi^2 v^{2}}\int_{-\infty}^{\infty}\frac{\mathrm{d}k}{k}\left(\frac{1}{2}+\frac{1}{e^{\beta k}-1}\right).
\end{equation}
This contribution is infrared divergent both at zero and at finite temperature, indicating $\langle{\rm e}^{i{\rm P}(X)}\rangle=0$, i.e., lack of phase coherence.

To demonstrate the absence of BEC, we further calculate the off-diagonal long-range order (ODLRO).  The gauge-invariant ODLRO of the system is given by
\begin{equation}
 \langle\Phi^*(X_{1})\Phi(X_{2}){\rm e}^{iq\int_{X_{1}}^{X_{2}}A_{\mu}dx^{\mu}}\rangle\propto\rho_{1}^{N}\rho_{2}^{N} {\rm e}^{-\frac{1}{4}qB(\rho_{1}^{2}+\rho_{2}^{2})-iN(\theta_{1}-\theta_{2})}\langle {\rm e}^{-i({\rm P}(X_{1})-{\rm P}(X_{2}))}\rangle.\label{gauge_invariant_ODLRO}
\end{equation}
Since there is no translational invariance in the $x-y$ plane, we analyze the decay of the ODLRO along the $z$-direction.  The correlation function of the phase factor can be evaluated as
\begin{equation}
\langle {\rm e}^{-i({\rm P}(X_{1})-{\rm P}(X_{2}))}\rangle=\frac{\int{\cal D}{\rm P}\left[{\rm e}^{-i{\rm P}(X_{1})}{\rm e}^{i{\rm P}(X_{2})}\right]{\rm e}^{S_{\rm P}}}{\int{\cal D}{\rm P}{\rm e}^{S_{\rm P}}}
={\rm e}^{-\frac{1}{2}\langle [{\rm P}(X_1)-{\rm P}(X_2)]^2\rangle}
\end{equation}
This quantity can be evaluated by using the fact
\begin{equation}
\langle [{\rm P}(X_1)-{\rm P}(X_2)]^2\rangle=G(X_{1}, X_{1})-2G(X_{1}, X_{2})+G(X_{2}, X_{2}),
\end{equation} 
For $\tau_1=\tau_2$, the Goldstone mode contribution ($s=l=0$) to the thermal propagator can be evaluated as
\begin{equation}
G(X_1,X_2)=\frac{qBT}{2\pi v^{2}}\sum_{n=-\infty}^{\infty}\int_{-\infty}^{\infty}\frac{\mathrm{d}k_{z}}{2\pi}\frac{\mathrm{e}^{-ik(z_{1}-z_{2})}}{\omega_{n}^{2}+k_{z}^{2}}
\end{equation}
Therefore, we obtain (see Appendix B)
\begin{equation}
    \begin{aligned}
       \langle [{\rm P}(X_1)-{\rm P}(X_2)]^2\rangle&=\frac{qB}{2\pi v^{2}}\int_{-\infty}^{\infty}\frac{{\rm d}k}{2\pi}\frac{1}{k}\left(1+\frac{2}{{\rm e}^{\beta k}-1}\right)\left[1-{\rm e}^{-ik(z_{1}-z_{2})}\right]\\
        &=\frac{qB}{2\pi^{2}v^{2}}\int_{0}^{\infty}\frac{\mathrm{d}k}{k}\left(1+\frac{2}{{\rm e}^{\beta k}-1}\right)\left\{1-\cos [k(z_{1}-z_{2})]\right\}\\
        &=\frac{qB}{2\pi^{2}v^{2}}\ln \left[\sinh\left(\frac{\pi |z_{1}-z_{2}|}{\beta}\right)\right]+\text{const}.
    \end{aligned}\label{ODLRO}
\end{equation}
For $|z_{1}-z_{2}|\to\infty$, the correlation function behaves as
\begin{equation}
\langle {\rm e}^{-i({\rm P}(X_{1})-{\rm P}(X_{2}))}\rangle\sim \exp{\left(-\frac{qBT}{4\pi v^2}|z_1-z_2|\right)}. \label{result}
\end{equation}
At $T\neq0$, it vanishes exponentially as $|z_{1}-z_{2}|\to\infty$, indicating the absence of ODLRO at any nonzero temperature. This result is in accordance to the Coleman-Mermin-Wagner-Hohenberg theorem \cite{CMWH}: The Goldstone boson is quasi-one-dimensional, and hence the formation of symmetry-breaking homogeneous long-range order along the $z$ direction is forbidden.

Beyond the discussions above, one possible concern is that, in principle, the $U(1)$ gauge field should also be treated as a dynamical degree of freedom. In this case, the gauge field could absorb the would-be Goldstone mode and become massive via the Anderson-Higgs mechanism. Since there is no physical Goldstone mode, our above conclusion appears invalid. However, this plausible concern does not influence the result. We emphasize here that the ground state $\Phi_{c}$ is inhomogeneous in the x-y plane, and hence there is no translational symmetry in the x-y plane. For this reason, the gauge field propagator is also quasi-one-dimensional. Qualitatively, the long-range behavior of the two-point correlation function of such a massive quasi-one-dimensional gauge field is
\begin{equation}
    \lim_{|z_{1}-z_{2}|\to\infty}\langle A_{3}(z_{1})A_{3}(z_{2})\rangle\sim Ke^{-M|z_{1}-z_{2}|},
\end{equation}
where $A_{3}$ is the z-component of the gauge field, $K$ is a constant added for dimensional consistency, and $M$ is the mass acquired through the Anderson-Higgs mechanism. The ODLRO between two points that differ only in their z-coordinate could be calculated similarly:
\begin{equation}
    \langle\Phi^*(X_{1})\Phi(X_{2}){\rm e}^{iq\int_{X_{1}}^{X_{2}}A_{\mu}dx^{\mu}}\rangle\propto \exp\Bigl[-\frac{1}{2}\int_{z_{1}}^{z_{2}}\mathrm{d}z\int_{z_{1}}^{z_{2}}\mathrm{d}z'\langle A_{3}(z)A_{3}(z')\rangle\Bigr]\propto\exp\Bigl[-\frac{K|z_{1}-z_{2}|}{2M}\Bigr].
\end{equation}
Then the result in Eq. (\ref{result}) is qualitatively reproduced, which agrees with our expectation that the ODLRO in Eq. (\ref{gauge_invariant_ODLRO}) is gauge-invariant. However, owing to the use of a rotating coordinate frame, the quantitative analytical computation of the gauge field correlation functions proves exceedingly challenging. Therefore, we keep the phase fluctuation as a dynamical field and integrate over it when calculating the ODLRO, similar to the standard paradigm for studying superconductivity in condensed matter theory.

\section{Summary}
Charged pion condensation driven by rotation in a parallel magnetic field was previously proposed by Liu and Zahed~\cite{PRL120:032001}. In this work, we have studied the critical temperature of charged pion condensation and have shown that it is absent. The charged pions are described by a complex scalar field with a quartic self-interaction.  We evaluate the partition function and the critical temperature using the standard field-theoretical method.
For non-interacting bosons, the critical temperature can only be determined for a system with fixed angular momentum.  We find that the critical temperature of the non-interacting system is zero due to the fact that the system
is quasi-one-dimensional, indicating that non-interacting bosons do not condense. For interacting bosons, we show that the order parameter is zero and the off-diagonal long-range order is absent at any nonzero temperature
because of the quasi-one-dimensional feature, in accordance with the Coleman-Mermin-Wagner-Hohenberg theorem.

{\bf Acknowledgments:} The work is supported by the National Natural Science Foundation of China (Grant No. 11775123).

%\section{Appendix}
\appendix

\section{Analysis of the effective potential}
For convenience, we denote the coefficients of the $v_{l}^{2}$ terms in (\ref{effective_potential}) as 
\begin{equation}
    \alpha_{l}=qB+m^{2}-(\mu+l\Omega)^{2}.
\end{equation}
To determine the minimum of the effective potential, we can focus primarily on the contributions of those modes with negative $\alpha_{l}$. The summation in (\ref{classical_solution}) is then truncated at $l^{*}$, where $l^{*}$ is the lower bound satisfying $\alpha_{l}<0$. For the sake of simplicity, we consider the case $\mu=0$. As $\Omega$ increases from zero, the first coefficient that becomes negative is $\alpha_{N}$, with the first critical angular speed
\begin{equation}
\Omega_{\rm c 1}=\frac{\sqrt{qB+m^{2}}}{N}.
\end{equation}
As $\Omega$ increases further, $\alpha_{N-1},\alpha_{N-2},\cdots$ will become negative sequentially, with critical angular speeds
\begin{equation}
\Omega_{\rm c 2}=\frac{\sqrt{qB+m^{2}}}{N-1},\ \ \ \Omega_{\rm c 3}=\frac{\sqrt{qB+m^{2}}}{N-2},\ \ \ \cdots
\end{equation}
(1) In the region $\Omega_{\rm c1}<\Omega<\Omega_{\rm c2}$, only the $l=N$ mode condenses, and we  write
\begin{equation}
\Phi_{\rm c}=v_Nf_N(\rho,\theta),\ \ \ \ \ \ \ f_l(\rho,\theta)\equiv\sqrt{\frac{1}{|l|!}}{\rm e}^{il\theta}\varphi_{0l}(\rho).
\end{equation}
The effective potential at tree level takes the form
\begin{equation}
 \mathcal{V}_{\text{eff}}^{(0)}(v_{N})=(qB+m^{2}-N^2\Omega^{2})v_{N}^{2}+\lambda C_{N,N,N,N}v_{N}^{4}.
\end{equation}
The phase transition is second-order since $C_{N,N,N,N}>0$. For instance, setting 
\begin{equation}
m^{2}=qB,\ \ \ \ \lambda=0.01,\ \ \ \ \Omega=\frac{\sqrt{2qB}}{N-0.5},
\end{equation}
the effective potential can be numerically evaluated as
\begin{equation}
\mathcal{V}_{\text{eff}}^{(0)}(v_{1})=-0.020151qBv_{N}^{2}+0.000281742v_{N}^{4}.
\end{equation}
The minimum is located at $v_{N}=5.98009\sqrt{qB}$.\\
(2) In the region $\Omega_{\rm c2}<\Omega<\Omega_{\rm c3}$, the $l=N$ and $l=N-1$ modes may condense, and we write
\begin{equation}
\Phi_{\rm c}=v_Nf_N(\rho,\theta)+v_{N-1}f_{N-1}(\rho,\theta).
\end{equation}
The effective potential at tree level takes the form
\begin{equation}
    \begin{aligned}
        \mathcal{V}_{\text{eff}}^{(0)}(v_{N},v_{N-1})&=(qB+m^{2}-N^2\Omega^{2})v_{N}^{2}+[qB+m^{2}-(N-1)^{2}\Omega^{2}]v_{N-1}^{2}\\
        &+\lambda(C_{N,N,N,N}v_{N}^{4}+C_{N-1,N-1,N-1,N-1}v_{N-1}^{4}+4C_{N,N,N-1,N-1}v_{N}^{2}v_{N-1}^{2}).
    \end{aligned}
\end{equation}
For instance, setting 
\begin{equation}
m^{2}=qB,\ \ \ \ \lambda=0.01, \ \ \ \ \Omega=\frac{\sqrt{2qB}}{N-1.5},
\end{equation}
the effective potential can be numerically evaluated as
\begin{equation}
    \begin{aligned}
        \mathcal{V}_{\text{eff}}^{(0)}(v_{N},v_{N-1})&=-0.0613775qBv_{N}^{2}-0.0203561qBv_{N-1}^{2}\\
        &+0.000281742v_{N}^{4}+0.000283158v_{N-1}^{4}+0.00112697v_{N}^{2}v_{N-1}^{2}.
    \end{aligned}
\end{equation}
The minimum is located at $v_{N}=10.4367\sqrt{qB},\ v_{N-1}=0$.\\
(3) In the region $\Omega_{\rm c3}<\Omega<\Omega_{\rm c4}$, the $l=N$,$l=N-1$ and $l=N-2$ modes may condense, and we write
\begin{equation}
\Phi_{\rm c}=v_Nf_N(\rho,\theta)+v_{N-1}f_{N-1}(\rho,\theta)+v_{N-2}f_{N-2}(\rho,\theta).
\end{equation}
The effective potential at tree level takes the form
\begin{equation}
    \begin{aligned}
        \mathcal{V}_{\text{eff}}^{(0)}(v_{N},v_{N-1},v_{N-1})&=(qB+m^2-N^2\Omega^{2})v_{N}^{2}+[qB+m^2-(N-1)^{2}\Omega^{2}]v_{N-1}^{2}\\
        &+[qB+m^{2}-(N-2)^{2}\Omega^{2}]v_{N-2}^{2}+\lambda\Bigl(C_{N,N,N,N}v_{N}^{4}+C_{N-1,N-1,N-1,N-1}v_{N-1}^{4}\\
        &+C_{N-2,N-2,N-2,N-2}v_{N-2}^{4}+4C_{N,N,N-1,N-1}v_{N}^{2}v_{N-1}^{2}+4C_{N,N,N-2,N-2}v_{N}^{2}v_{N-2}^{2}\\
        &+4C_{N-1,N-1,N-2,N-2}v_{N-1}^{2}v_{N-2}^{2}+4C_{N,N-1,N-1,N-2}v_{N}v_{N-1}^{2}v_{N-2}\Bigr).
    \end{aligned}
\end{equation}
For instance, setting 
\begin{equation}
m^{2}=qB,\ \ \ \ \lambda=0.01, \ \ \ \ \Omega=\frac{\sqrt{2qB}}{N-2.5},
\end{equation}
the effective potential can be numerically evaluated as
\begin{equation}
    \begin{aligned}
        \mathcal{V}_{\text{eff}}^{(0)}(v_{N},v_{N-1},v_{N-2})&=-0.103879qBv_{N}^{2}-0.0620118qBv_{N-1}^{2}-0.0205654qBv_{N-2}^{2}\\
        &+0.000281742v_{N}^{4}+0.000283158v_{N-1}^{4}+0.000284596v_{N-2}^{4}\\
        &+0.00112697v_{N}^{2}v_{N-1}^{2}+0.00112131v_{N}^{2}v_{N-2}^{2}+0.00113263v_{N-1}^{2}v_{N-2}^{2}\\
        &+0.00112696v_{N}v_{N-1}^{2}v_{N-2}.
    \end{aligned}
\end{equation}
Numerical minimization shows that the minimum is located at $v_{N}=13.5776\sqrt{qB}$, while $v_{N-1}$ and $v_{N-2}$ are smaller than $v_{N}$ by more than a dozen orders of magnitude. This is consistent with the previous result \cite{PRD106:094010}. A reasonable conjecture is that, if we consider all the $l>0$ modes, a self-consistent variational calculation will give the result $v_N\neq0$ and all other $v_l$s are vanishingly small. 

A more recent study \cite{PhysRevD.109.056024} indicates that in some parameter regions, the system might have more than one $v_{l}$ is non-negligible, and such a ground state is called "vortex lattice". An exact analytical treatment of this situation is exceedingly difficult. Nevertheless, the following argument can be made: the vortex lattice breaks the translational symmetry of the ground state more strongly than a single vortex, so the associated Goldstone mode and gauge field propagator remain quasi-one-dimensional. Therefore, our conclusion remains valid.

\section{Calculation of the integral in (\ref{ODLRO}) }

In (\ref{ODLRO}), we need to evaluate the integral 
\begin{equation}
C(z_{1}-z_{2})=\int_{0}^{\infty}\frac{\mathrm{d}k}{k}\left(1+\frac{2}{{\rm e}^{\beta k}-1}\right)\Big\{1-\cos [k(z_{1}-z_{2})]\Big\}.
\end{equation} 
We decompose it into a zero-temperature part $C_{0}(z_{1}-z_{2})$ and a finite temperature part $C_{T}(z_{1}-z_{2})$,
\begin{equation}
    \begin{aligned}
        C(z_{1}-z_{2})&=C_{0}(z_{1}-z_{2})+C_{T}(z_{1}-z_{2}),\\
        C_{0}(z_{1}-z_{2})&=\int_{0}^{\infty}\frac{\mathrm{d}k}{k}\Big\{1-\cos [k(z_{1}-z_{2})]\Big\},\\
        C_{T}(z_{1}-z_{2})&=\int_{0}^{\infty}\frac{\mathrm{d}k}{k}\frac{2}{{\rm e}^{\beta k}-1}\Big\{1-\cos [k(z_{1}-z_{2})]\Big\}.
    \end{aligned}
\end{equation}
For $C_{0}(z_{1}-z_{2})$, after introducing an ultraviolet cutoff $\Lambda$, it can be evaluated as
\begin{equation}
    C_{0}(z_{1}-z_{2})=\ln |z_{1}-z_{2}|+\ln\Lambda+\gamma_{\rm E},
\end{equation}
where $\gamma_{\rm E}$ is the Euler constant. For $C_{T}(z_{1}-z_{2})$, we use the expansion of the Bose-Einstein distribution to obtain
\begin{equation}
    C_{T}(z_{1}-z_{2})=2\sum_{n=1}^{\infty}\int_{0}^{\infty}\frac{\mathrm{d}k}{k}{\rm e}^{-n\beta k}\Big\{1-\cos [k(z_{1}-z_{2})]\Big\}.
\end{equation}
For each $n$, the integral can be evaluated as
\begin{equation}
    I_{n}(z_{1}-z_{2})=\int_{0}^{\infty}\frac{\mathrm{d}k}{k}{\rm e}^{-n\beta k}\Big\{1-\cos [k(z_{1}-z_{2})]\Big\}=\frac{1}{2}\ln\left[1+\frac{(z_{1}-z_{2})^{2}}{\beta^{2}n^{2}}\right].
\end{equation}
The summation over $n$ can be completed by using the identity 
\begin{equation}
 \prod_{n=1}^{\infty}\left(1+\frac{z^{2}}{n^{2}\pi^{2}}\right)=\frac{\sinh z}{z}.
\end{equation}
We obtain
\begin{equation}
    C_{T}(z_{1}-z_{2})=\ln\left[\frac{\sinh (\pi T|z_{1}-z_{2}|)}{\pi T|z_{1}-z_{2}|}\right].
\end{equation}
Then we add the results for $C_{0}(z_{1}-z_{2})$ and $C_{T}(z_{1}-z_{2})$ to obtain the result in (\ref{ODLRO}).

\bibliographystyle{apsrev4-1}

\end{document}